\documentclass[square]{ws-procs975x65}

\usepackage{graphicx}

\newcommand{\tens}[1]{{\boldsymbol{#1}}}

\newcommand{\n}[1]{\label{#1}}
\newcommand{\be}{\begin{equation}}
\newcommand{\ee}{\end{equation}}
\newcommand{\ba}{\begin{eqnarray}}
\newcommand{\ea}{\end{eqnarray}}

\newcommand{\beq}{\begin{equation}}
\newcommand{\eeq}{\end{equation}}
\newcommand{\beqa}{\begin{eqnarray}}
\newcommand{\eeqa}{\end{eqnarray}}

\newcommand{\hook}{\raisebox{-0.35ex}{\makebox[0.6em][r]
{\scriptsize $-$}}\hspace{-0.15em}\raisebox{0.25ex}{\makebox[0.4em][l]{\tiny
 $|$}}}

\begin{document}

\title{Black hole spacetimes with Killing--Yano symmetries}

\author{David Kubiz\v n\'ak}

\address{DAMTP, University of Cambridge,\\
Wilberforce Road, Cambridge CB3 0WA, UK\\
E-mail: D.Kubiznak@damtp.cam.ac.uk\\
www.damtp.cam.ac.uk/people/d.kubiznak}

\begin{abstract}
We present a brief overview of black hole spacetimes admitting Killing--Yano tensors.
In vacuum these include Kerr-NUT-(A)dS metrics and certain black brane solutions.
In the presence of matter fields, (conformal) Killing--Yano symmetries are known 
to exist 
for the Plebanski--Demianski solution and (trivially) for any spacetime with spherical symmetry.
Special attention is devoted to generalized Killing--Yano tensors of black holes in minimal gauged supergravity. Several aspects directly related to the existence of 
Killing--Yano tensors---such as the Kerr--Schild form, algebraic type of spacetimes, and separability of field 
equations---are also briefly discussed.   
\end{abstract}


\bodymatter

\section{Introduction}\label{aba:sec1}

It is now more than forty years since hidden symmetries have been recognized to play an 
important role in black hole physics. The foundation of the subject dates back to 1968 and the seminal paper of Brandon Carter \cite{Carter:1968pr} who discovered a mysterious and unexpected constant of motion for geodesic trajectories in the rotating black hole spacetime of the Kerr geometry \cite{Kerr:1963}. In the following years it was recognized that the Carter's constant can be traced to the existence of hidden symmetries described by the {\em Killing} \cite{Stackel:1895, WalkerPenrose:1970} and {\em Killing--Yano} (KY) tensors \cite{Yano:1952, Penrose:1973, Floyd:1973}. Subsequent studies have also shown that these symmetries are deeply connected with complete integrability of geodesics motion, separability of fundamental field equations and corresponding symmetry operators,  various conserved quantities, worldline supersymmetry of spinning particles, and many other exceptional properties of the Kerr, and, more generally, Kerr-NUT-(A)dS \cite{Carter:1968pl, Carter:1968cmp} spacetimes
(for recent review see \cite{FrolovKubiznak:2008}). 

With 
motivations from 
string theory and braneworld scenarios, hidden symmetries and spacetimes with hidden symmetries were recently studied in higher dimensions. It turns out,
that vacuum (including a cosmological constant) spacetimes admitting KY tensors include physically interesting geometries such as the most general known higher-dimensional  Kerr-NUT-(A)dS metrics
\cite{ChenEtal:2006cqg}. As a consequence, the properties of these (spherical) black holes are  very similar to the properties of their four-dimensional counterparts
\cite{FrolovKubiznak:2008, Kubiznak:phd, Frolov:2009}. 
Another class of spacetimes which possesses hidden symmetries is a class of black branes
where the KY symmetry is inherited from the black hole base space. On the other hand,
the existence of a non-degenerate CKY 2-form restricts the spacetime to a special algebraic type D \cite{MasonTaghavi:2008}. This implies that (non-degenerate) KY tensors cannot exist for 5D non-spherical black holes, such as black rings or black saturns.
An interesting question is whether hidden symmetries can be found for black holes in the presence of matter fields, such as Kaluza--Klein black holes or black holes of various supergravities.
Although (conformal) Killing tensors are known for a wide class of supergravity black holes, see, e.g., \cite{Chow:2008} and references therein, this is not the case for KY tensors.
An example of spacetime admitting a KY symmetry is the Kerr-Newman black hole, which may be viewed as a solution of $D=4$ supergravity. A proper generalization of this symmetry was recently found for black holes of $D=5$ minimal gauged supergravity. For black holes of other supergravities KY tensors are (if they exist) yet to be discovered. 

In this paper, we try to briefly overview the picture of recent progress in the field of hidden symmetries. 
As (at least for the vacuum case) there exist several extended reviews on the subject, see, e.g, \cite{FrolovKubiznak:2008, Kubiznak:phd, Frolov:2009},
we mainly concentrate on two things: i) we overview black hole spacetimes for which hidden symmetries are known to exist ii) we summarize certain aspects related to generalized KY tensors in minimal gauged supergravity. 
Throughout the paper we follow notations of \cite{FrolovKubiznak:2008}.

\section{Vacuum spacetimes with hidden symmetries}
\subsection{Black holes}
In vacuum $D$-dimensional (rotating) black hole spacetimes the central hidden symmetry is associated with a 
{\em closed conformal Killing--Yano} (CCKY) 2-form \cite{FrolovKubiznak:2007, KubiznakFrolov:2007}, which is a special case of a
{\em conformal Killing--Yano} (CKY) tensor introduced by Kashiwada and Tachibana
\cite{Kashiwada:1968, Tachibana:1969}.
The CKY tensor $\tens{k}$ of rank $p$ is a $p$-form which for any vector field $\tens{X}$
obeys (the Hodge star invariant) equation 
\be\n{CKY}
{\nabla}_{X} \tens{k}={1\over p+1} \tens{X}\hook \tens{d}\tens{k}
-{1\over D-p+1}\tens{X}^{\flat}\wedge
\tens{\delta} \tens{k}\, .
\ee 
If the first term on the r.h.s. vanishes one has a CCKY tensor. The vanishing of the second term means that one is dealing with a KY tensor.
These `dual' special cases (the Hodge star transforms a CCKY tensor into a KY tensor and vice versa) are of particular importance. They give rise to Killing tensors and associated conserved quantities. 
Moreover, the CCKY tensors form an algebra with respect to the wedge product; 
a wedge product of two CCKY tensors is again a CCKY tensor  \cite{KrtousEtal:2007jhep}.

The most general metric element admitting a CCKY 2-form $\tens{h}$, 
\be\label{PCKY}
\nabla_c h_{ab}=2g_{c[a}\xi_{b]}\,,\quad 
{\xi}_a={1\over D-1}\,\nabla_c h^{c}_{\ a}\, ,
\ee
 was constructed 
by Houri {\em et al.} \cite{HouriEtal:2008b, HouriEtal:2009a}.
Depending on the character of eigenvalues of $\tens{h}^2$, such a metric consists of three parts: i)~a `canonical Kerr-NUT-(A)dS spacetime' \cite{HouriEtal:2007, KrtousEtal:2008} which corresponds to functionally independent eigenvalues ii)~the (unspecified) K\"ahler 
manifolds---corresponding to nonzero constant eigenvalues and iii)~an `arbitrary metric' on a space of zero eigenvalue.
Specifically, when all the eigenvalues are functionally independent 
$\tens{h}$ was called a {\em principal conformal Killing--Yano} (PCKY) tensor \cite{FrolovKubiznak:2007, KubiznakFrolov:2007, KrtousEtal:2007jhep} and the metric possesses many interesting properties. Namely, since the PCKY tensor is completely non-degenerate, one can extract from it a sufficient number of explicit and hidden symmetries \cite{KrtousEtal:2007jhep} which ensure complete integrability of geodesic motion \cite{PageEtal:2007, KrtousEtal:2007prd, HouriEtal:2008a},
separability of the Hamilton--Jacobi \cite{FrolovEtal:2007}, Klein--Gordon \cite{FrolovEtal:2007, SergyeyevKrtous:2008}, Dirac \cite{OotaYasui:2008, Wu:2008, Wu:2008b}, and stationary string \cite{KubiznakFrolov:2008} equations, as well as separability of certain gravitational \cite{KunduriEtal:2006, OotaYasui:2009} perturbations. The metric is of the algebraic type D \cite{HamamotoEtal:2007}. The problem of parallel transport reduces to a set of first order differential equations \cite{ConnellEtal:2008, KubiznakEtal:2009a}.
When the vacuum Einstein equations are imposed the metric describes the most general known Kerr-NUT-(A)dS black hole spacetime \cite{ChenEtal:2006cqg}.  
 Performing a certain limit in which the PCKY tensor becomes completely degenerate one can obtain \cite{Kubiznak:2009} the most general known Einstein--K\"ahler and Einstein--Sasaki manifolds \cite{ChenEtal:2006cqg, ApostolovEtal:2004}. 
The (Wick-rotated) metric $\tens{g}$, together with its PCKY tensor $\tens{h}=\tens{db}$, can be written in the following multi Kerr--Schild form \cite{ChenLu:2008} ($n=[D/2], \varepsilon=D-2n$):\footnote{The expressions for $\tens{g}$ and $\tens{h}$ in the so called Darboux basis can be found, e.g., in reviews \cite{FrolovKubiznak:2008, Kubiznak:phd, Frolov:2009}. 
}
\be\label{KerrSchild}
\tens{g}=\tens{g}_{\mbox{\tiny (A)dS}}+\sum_{\mu=1}^n\frac{2b_\mu x_\mu^{1-\varepsilon\!\!\!\!}}{U_\mu}\,\tens{l}_\mu^2\,, \qquad
\tens{h}=\sum_{\mu=1}^n x_\mu \tens{d}x_\mu\wedge \tens{l}_\mu\,,
\ee
where parameters $b_\mu$ stand for mass and NUT charges, and the (A)dS metric $\tens{g}_{\mbox{\tiny (A)dS}}$ is written as 
\ba
\tens{g}_{\mbox{\tiny (A)dS}}\!&=&\!-\sum_{\mu=1}^n\Bigl(\frac{X_\mu}{U_\mu}\, \tens{l}_\mu^2 -\tens{l}_\mu \tens{d}x_\mu-\tens{d}x_\mu\tens{l}_\mu \Bigr)
+\frac{\varepsilon c}{A^{(n)}}\,\tens{l}_0^2\,,\quad 
X_{\mu}=\sum\limits_{k=\varepsilon}^{n}c_kx_{\mu}^{2k}+\frac{\varepsilon c}{x_{\mu}^2}\,,\nonumber\\
\tens{l}_\mu\!&=&\!\sum_{k=0}^{n-1}A_\mu^{(k)}\tens{d}\psi_k\,,\quad  
\tens{l}_0=\sum_{k=0}^n A^{(k)}\tens{d}\psi_k\,,\quad 
\tens{b}=\frac{1}{2}\sum_{k=0}^{n-1}A^{(k+1)}\tens{d}\psi_k\,,\\
U_{\mu}\!&=&\!\prod_{\nu\ne\mu}(x_{\nu}^2-x_{\mu}^2)\,,\quad
A^{(k)}_\mu=\!\!\!\sum_{\substack{\nu_1<\dots<\nu_k\\\nu_i\ne\mu}}\!\!\!\!\!x^2_{\nu_1}\dots x^2_{\nu_k}\,,\quad 
A^{(k)}=\!\!\!\!\!\sum_{\nu_1<\dots<\nu_k}\!\!\!\!\!x^2_{\nu_1}\dots x^2_{\nu_k}\,.\nonumber
\ea
Parameter $c_n$ is proportional to the cosmological constant, other parameters 
$c_k, c$ are related to rotations; metric functions 
$X_\mu(x_\mu)$ are linear in all of them. 
Forms $\tens{l}_\mu$ and $\tens{l}_0$ are 
nonzero(zero)-eigenvalue eigenforms of $\tens{h}$ and correspond to principal null directions (WANDs) of the metric \cite{HamamotoEtal:2007, KrtousEtal:2008, MasonTaghavi:2008, Kubiznak:phd}. The expression \eqref{KerrSchild} nicely demonstrates the connection between the existence of the PCKY tensor and the form of the corresponding Kerr--Schild structure. (For a more general discussion on higher-dimensional Kerr-Schild 
spacetimes see \cite{OrtaggioEtal:2009a}). 

\subsection{Black branes}
Another class of higher-dimensional spacetimes which possess KY symmetries and do not necessary fall into the class described above are black branes with the base that allows a KY tensor. Let $\tens{g}_B$ be a (black hole) base space admitting a KY tensor $\tens{f}$ (CKY is not enough). Then, a $p$-brane 
\be
\tens{g}=\tens{g}_B+\sum_{a=1}^p \tens{d}x^a\tens{d}x^a\,,
\ee 
possesses the same KY tensor $\tens{f}$ (now understood as a higher-dimensional one).
An obvious example of this is a Kerr-string which inherits a KY 2-form from the Kerr spacetime. 
Slightly more generally, let $\tens{g}_B$ be a base space admitting a KY $q$-form $\tens{f}_B$, then $\tens{g}(x,y)=h^2(x)\tens{g}_B(y)+\gamma_{ab}(x)\tens{d}x^a\tens{d}x^b$ inherits a KY tensor $\tens{f}=h^{q+1}(x)\tens{f}_B$. In particular, a product spacetime inherits all the KY tensors of individual constituents. See also metric \eqref{ss} below.

\section{Non-vacuum spacetimes with hidden symmetries}
As mentioned in the introduction, there are only few known examples of non-vacuum black hole spacetimes admitting (conformal) KY symmetries. 
In 4D the most important of them is a (7-parametric) family of type D solutions of the Einstein--Maxwell equations obtained by Plebanski and Demianski \cite{PlebanskiDemianski:1976}.
Among others, this metric element describes 
the Kerr--Newman black hole, the various generalizations of the C-metric, or even the, recently discovered, black hole spacetimes with a conformally coupled scalar field 
\cite{AnabalonMaeda:2009, CharmousisEtal:2009}.
The metric admits a CKY 2-form, which becomes a KY 2-form when the `acceleration' parameter is removed.\footnote{%
In fact, after the acceleration is removed, the metric element attains a form of the  four-dimensional canonical Kerr-NUT-(A)dS spacetime described in Sec. 2.
Such an element was already obtained by Carter in 1968 \cite{Carter:1968cmp} as the most general metric admitting separability of the Hamilton--Jacobi and Klein--Gordon equations.
} 
Explicit expressions for the 
(conformal) KY tensors of the general metric and its various subclasses can be found in  \cite{KubiznakKrtous:2007}.
Another example of spacetime admitting CKY 2-form is a dilatonic C-metric obtained recently by 
a KK-reduction of single spinning black ring \cite{Durkee:2009}.

Different class admitting KY symmetries are spherically symmetric spacetimes.
KY tensors there are `inherited from the sphere'. Namely, a `generalized' spherically symmetric spacetime  
\be\label{ss}
\tens{g}=\gamma_{ab}(x)\tens{d}x^a\tens{d}x^b+h^2(x)\tens{d}\omega^2_{d}\,,
\ee
where $\tens{d}\omega_{d}^2$ denotes the metric induced on a unit
$d$-sphere (parameterized by angles $\theta_i\in[0,\pi]$, $i=1,\dots, d-1$ and $\varphi\in[0,2\pi]$)
\be
\tens{d}\omega^2_d=\sum_{i=1}^{d-1}\tens{e}_i^2+\tens{e}_\varphi^2\,,\quad
\tens{e}_i=\prod_{j<i}\sin\theta_j\,\tens{d}\theta_{\!i}\,,\ \ 
\tens{e}_\varphi=\prod_{j=1}^{d-1}\sin\theta_{\!j} \,\tens{d}\varphi\,,
\ee
possesses the following ($i=1,\dots, d-1$) KY tensors of decreasing rank:\footnote{%
It should be stressed that tensors $\tens{f}_{i}^{(\omega_d)}$ are not the only KY forms on 
a unit sphere.
In fact, $S^d$ admits a maximum possible number, $n=(d+1)!/[(d-p)!(p+1)!]$, of independent 
KY tensors of each rank $p$, which coincide with the coclosed minimal eigenvalue eigenforms of the Laplace operator \cite{Semmelmann:2002}. 
}
\be
\tens{f}_{i}=h^{d-i+2}(x)\tens{f}_{i}^{(\omega_d)}\,,\quad \tens{f}_{i}^{(\omega_d)}=
\bigl(\prod_{j<i}\sin\theta_{\!j}\bigr)\,
\tens{e}_i\wedge \tens{e}_{i+1}\wedge\dots\wedge  
\tens{e}_{d-1}\wedge \tens{e}_{\varphi}\,.
\ee
In particular, this applies to spherically symmetric black holes of a-model \cite{Ortin:2004}, or to charged black strings (see, e.g., \cite{FrolovShoom:2009} and references therein).

\section{Generalized KY symmetries: Black hole of minimal gauged supergravity}
When dealing with CKY symmetries defined by Eq. \eqref{CKY}, there is an important limitation which restricts the spacetime to a spacetime of special algebraic type.
Namely, a spacetime admitting a (non-degenerate) CKY 2-form must be of the special algebraic type D \cite{Collinson:1974, Stephani:1978, GlassKress:1999, MasonTaghavi:2008}. Such a restriction a priori excludes most of the black holes with matter fields, such as those of various supergravities.
On the other hand, some of these black holes are known to admit Killing tensors
and allow separability of the Hamilton--Jacobi and Klein--Gordon equations,
see, e.g., \cite{Chow:2008} and references therein. An interesting question is whether it is possible to generalize the concept of KY symmetries for these solutions, while preserving most of its fundamental physical properties and implications. The following example demonstrates that 
a natural generalization exists for a black hole of minimal gauged supergravity.

It was shown by Wu \cite{Wu:2009a} that the most general known black hole solution of $D=5$
minimal gauged supergravity \cite{ChongEtal:2005b} (which is a solution of the Einstein--Maxwell theory with a specific Chern--Simons coupling) admits a {\em generalized PCKY} 2-form $\tens{h}$.
Such a 2-form obeys a modified Eq. \eqref{PCKY}, where an additional term, proportional to the electromagnetic flux $\tens{*F}$, is present \cite{Wu:2009a, KubiznakEtal:2009b},
\be\label{PCKY2}
\nabla_c h_{ab}=2g_{c[a}\xi_{b]}-\frac{1}{\sqrt{3}}\,(*F)_{cd[a}h^d_{\ \,b]}\,.
\ee
This generalization can be naturally described \cite{KubiznakEtal:2009b} by the original Eq. \eqref{PCKY} if all the derivatives therein 
are understood as in the presence of a torsion field $\tens{T}=\tens{*F}/\sqrt{3}$. 
Such a torsion is `harmonic', $\tens{\delta}^T\tens{T}=0=\tens{d}^T\tens{T}$, due to the Maxwell's equations, and the identification has many appealing features. 
CKY tensors in the presence of torsion 
preserve most of the properties of `standard' CKY tensors. In particular, the generalized PCKY tensor $\tens{h}$ generates a Killing tensor and implies the existence of symmetry operators for the Hamilton--Jacobi, Klein--Gordon, and torsion-modified Dirac equations. In the background  \cite{ChongEtal:2005b} these equations and their charged versions were separated in \cite{DavisEtal:2005, Wu:2009a,  Wu:2009b}, stationary string equations for equal angular momenta
were separated in \cite{AhmedovAliev:2008}. 
The (Wick-rotated) metric $\tens{g}$ and the electromagnetic flux $\tens{F}=\tens{dA}$ are given by \cite{ChongEtal:2005b, LuEtal:2008b, KubiznakEtal:2009b}
\ba\label{charged}
\tens{g}&=&{\frac{x-y}{4X}}\,\tens{d}x^2-
{\frac{x-y}{4Y}}\,\tens{d}y^2-
\frac{X(\tens{d}t+y\tens{d}\phi)^{2\!}}{{x(x-y)}}
+\frac{{Y(\tens{d}t+x\tens{d}\phi)^{2\!}}}{{y(x-y)}}-
\frac{\mu^2}{{xy}}\,\tens{l}_0^2\,,\nonumber\\
\tens{A}&=&\frac{\sqrt{3}\mu}{x-y}\,\bigl[q(\tens{d}t+y\tens{d}\phi)-p(\tens{d}t+x\tens{d}\phi)\bigr]\,,
\ea
where  
\ba
\tens{l}_0&=&\tens{d}t\!+\!(x\!+\!y)\tens{d}\phi\!+\!xy\tens{d}\psi-\frac{qy(\tens{d}t+y\tens{d}\phi)}{x-y}+\frac{px(\tens{d}t+x\tens{d}\phi)}{x-y}\,,\nonumber \\  
X&=&\mu^2(1+q)^2+ax+cx^2+\frac{1}{12}\Lambda x^3\,,\quad 
Y=\mu^2(1+p)^2+by+cy^2+\frac{1}{12}\Lambda y^3\,\label{XYcharged}.
\ea
The generalized PCKY tensor $\tens{h}$ in this background is closed and can be generated from a potential, $\tens{h}=\tens{db}$, with   
$\tens{b}=-\frac{1}{2}\bigl[(x+y)\tens{d}t+xy\tens{d}\phi\bigr]$.
The metric is of type $I_i$ and the eigenvectors of $\tens{h}$ are principal 
null directions. 
Similar to the vacuum case, all the Killing fields can be `derived' from the knowledge of $\tens{h}$
\be\label{KV}
\xi^a_0=\xi^a=(\partial_{t})^a\,,\quad
\xi_1^a=K^{a}_{\ b}\xi^b+\frac{1}{4\sqrt{3}}(*F)^{abc}K_{bd}h^d_{\ c}=(\partial_\phi)^a\,,\quad
\xi_2^a=\frac{\mu}{2}(*{h}^{\wedge 2})^a=(\partial_{\psi})^a\,. 
\ee
Here, $K_{ab}=(*h)_{acd}(*h)_b^{\ cd}=h_{ac}h_b^{\ c}-\frac{1}{2}g_{ab}h^2\,,$ is a Killing tensor associated with $\tens{h}$. The construction \eqref{KV} is a `natural' generalization of the construction known from vacuum \cite{KrtousEtal:2007jhep} and 
underlines the following {uniqueness} result \cite{AhmedovAliev:2009}: the spacetime  \eqref{charged}--\eqref{XYcharged} is the most general spacetime with `harmonic' torsion field which admits a closed generalized PCKY tensor.
  
The motion of charged test particles is governed by the charged Hamilton--Jacobi equation
\be
\frac{\partial S}{\partial \lambda}+g^{ab}(\partial_a S+eA_a)(\partial_bS+eA_b)=0\,.
\ee 
The separability of this equation in the background \eqref{charged} follows from the separability of the complex Klein--Gordon field \cite{Wu:2009b} and the geometric optics approximation. 
This results in the following formula for the particle's momentum $\tens{p}$ (obeying $\dot p_a=eF_{ab}p^b$): 
\be
\tens{p}=\bigl(E-e\sqrt{3}\mu\frac{q\!-\!p}{x-y}\bigr)\tens{d}t+
\bigl(\Phi-e\sqrt{3}\mu\frac{qy\!-\!px}{x-y}\bigr)\tens{d}\phi+\Psi\tens{d}\psi+
\sigma_x\sqrt{\frac{W_x^2}{4xX^2}+\!\frac{V_x}{4X}}\,\tens{d}x+
\sigma_y\sqrt{\frac{W_y^2}{4yY^2}+\!\frac{V_y}{4Y}}\,\tens{d}y\,,
\ee
where $\sigma_x, \sigma_y=\pm$ are independent signs,
\ba
V_x\!&=&\!\kappa_0 x-\kappa_1-\frac{\Psi^2}{\mu^2x}\,,\ \ 
V_y=\kappa_0 y-\kappa_1-\frac{\Psi^2}{\mu^2y}\,,\ \ \kappa_0=p_ap^a\,,\nonumber\\
W_x\!&=&\!Ex^2-\Phi x+\Psi+q(\Psi-\sqrt{3}\mu ex)\,,\ \ 
W_y=Ey^2-\Phi y+\Psi+p(\Psi-\sqrt{3}\mu ey)\,,
\ea
the parameters $E, \Phi, \Psi$ are separation constants corresponding to 
the Killing fields \eqref{KV}, and $\kappa_1$ is a separation constant associated with the  Killing tensor $\tens{K}$. 
Let us finally mention that metric \eqref{charged} can be cast in a certain `generalized' Kerr--Schild form \cite{AlievCiftci:2009}. 
This form is, however,
fundamentally different to \eqref{KerrSchild}; in addition to standard Kerr--Schild terms 
(proportional to the squares of principal null directions) 
there appear additional cross terms of principal null directions with extra space-like
vector. 
 

\end{document}